\documentstyle[12pt,a4]{article}
\date{}

\begin{document}

\title{Electron conduction within Landau level tails of
medium-mobility GaAs / AlGaAs heterostructures}

\author{P. Svoboda$^a$, G. Nachtwei$^{b,d}$, C. Breitlow$^c$, S.
Heide$^d$ and M. Cukr$^a$ \\ \\ \\
\it $^a$ Institute of Physics, Academy of Sciences of the Czech
Republic \\ \it Cukrovarnick\'{a} 10 , CZ-162 00 Prague 6, Czech
Republic\\ \\
\it $^b$ Max-Planck-Institut f\"{u}r Festk\"{o}rperforschung \\
\it Heisenbergstr. 1 , D-70569 Stuttgart , Germany \\ \\
\it $^c$ Physikalisch-Technische Bundesanstalt Berlin \\
\it F\"{u}rstenwalder Damm 388 , D-12587 Berlin , Germany \\ \\
\it $^d$ Humboldt-Universit\"{a}t zu Berlin, AG Magnetotransport \\
\it Invalidenstr. 110 , D-10115 Berlin, Germany}

\maketitle

\vspace{20mm}

{\it Short title :} Conduction within Landau level tails

\vspace{10mm}

{\it PACS :} 7150, 7280E, 7340H

\vspace{20mm}

\hspace*{90mm} Typeset using \LaTeX

\newpage

\begin{abstract}
\noindent The temperature dependence of both components of the
resistivity tensor $\varrho_{xx}(T)$ and $\varrho_{xy}(T)$
has been studied at T $\geq$ 4.2 K
within IQHE plateaux around filling factors $\nu$=2 and $\nu$=4 of
medium-mobility GaAs/AlGaAs heterostructures. In the middle
of the mobility gap standard activated conductivity has been
found with activation energies $\Delta $ scaling well with $\hbar
\omega_{c} / 2$ .
 At filling factors slightly below $\nu$=2  another contribution adds to
 the activated conductivity at T $\leq$ 12 K. This additional contribution
 can be further enhanced at higher mesuring d.c. currents.
We suggest, that it arises due to enhanced electric field assisted
tunneling across potential barriers separating localized states
within the bulk of the sample
This effect contributes to the backscattering across the
sample leading to an enhanced longitudinal
conductivity. The additional contribution
to $\sigma_{xx}(T)$ can be reasonably well fitted to the formula for
the variable range hopping in strong magnetic fields indicating that
the hopping can persist even at temperatures well above 4.2K.
\end{abstract}

\newpage

\section{Introduction}
It is generally accepted that the integer quantum Hall effect
(IQHE) can be understood in terms of localized and extended
single electron states. While the latter form very narrow bands of a
width $\Gamma$ centered at Landau level energies $E_{N}$, the
former fill the mobility gap of a width ($\hbar \omega_{c} -
\Gamma$) and do not contribute to the longitudinal conductivity
$\sigma_{xx}$ at least in the limit T$\rightarrow$0.

Such a picture has been supported by the measurement of thermally
activated electron transport in 2DEG structures in strong
magnetic fields corresponding to an integer filling factor $\nu$ ,
i.e. to the center of a mobility gap. It has been found that
there is a finite temperature interval  $\Delta T$, where the
longitudinal conductivity $\sigma_{xx}(T)$ can be described by
the Arrhenius law

\begin{equation}\label{aktiv}
\sigma_{xx}(T) = \sigma_{xx}^o  e^{-\Delta/k_{B} T}
\end{equation}

\noindent with the activation energy $\Delta \cong \hbar \omega_c / 2$
indicating that the mobility edge $E_m$ virtually coincides with $E_N$
($\Gamma \rightarrow 0$).

The expression (\ref{aktiv})
appears as a special form of the general expression for the
temperature dependent conductivity $\sigma_{xx}(T)$ that reads

\begin{equation}\label{tepzav}
\sigma_{xx}(T) = \int \sigma_o(E) \frac{\partial f}{\partial E}
dE
\end{equation}

\noindent provided that
the Fermi - Dirac distribution $f(E)$ can be approximated
by the Boltzmann statistics. Since only the extended states at
$E \cong E_N$
contribute to $\sigma_o(E)$, the approximation (\ref{aktiv}) should
hold for $\mid E_F - E_N \mid >> k_BT$. If the Fermi energy $E_{F}$ lies
in the middle of a sufficiently wide mobility gap (i.e.
at $\nu$=N with N being a small even integer) this condition can
hold even at T $\geq$ 10K.

It has been found from the investigation of
high-mobility heterostructures GaAs/AlGaAs \cite{Clark}, that the
the prefactor in (\ref{aktiv}) reaches a universal value
$\sigma_{xx}^o \cong e^2/h$ independent on the sample and the
filling factor.
Although good fits to (\ref{aktiv}) in a finite temperature
interval have been reported for samples of widely different
parameters \cite{Clark,Usher,Weiss,Katayama}, the universality
of the prefactor $\sigma_{xx}^o$ has been disputed both
experimentally \cite{Katayama} and theoretically \cite{Polyak1}.
Recently, Polyakov and Shklovskii \cite{Polyak2} have
shown, that a universal temperature independent prefactor
in (\ref{aktiv}) can be derived from the
percolation theory in the limit of a long range impurity
potential, which is typical for high-mobility modulation doped
GaAs/AlGaAs heterostructures. They predict, that just in the
center of a mobility gap this prefactor equals to $2 e^2/h$ i.e.
it is twice as large as that found experimentally by  Clark
\cite{Clark}.

Deviations from a linear Arrhenius graph
(\ref{aktiv}) occur both at high and
low temperatures. The high temperature downward curvature has
been attributed either to a violation of the Boltzmann statistics
or to the
situation, where the electron mean free path becomes larger than
the perimeter $p_{T}$ of the percolation loops \cite{Polyak2}.
At sufficiently low temperatures an upward curvature is usually
observed and three basic explanations have been suggested. First,
variable range hopping (VRH) among localized states at $E_{F}$ is
expected to compete with the activated conduction at lowest
temperatures \cite{Polyak1,Ebert}. Non-ideal potential contacts
\cite{Komiyama} and a finite width W of broadened Landau levels
\cite{Usher} can both cause the observed low temperature
deviations from the simple activated conductivity described by
(\ref{aktiv}).

In this paper we investigate the temperature dependent
conductivity $\sigma_{xx}(T)$ beyond the breakdown of the IQHE
in the vicinity of the lowest even
filling factors as a function of the position within the
mobility  gap and of the measuring d.c. current $I$. The samples
studied were two modulation doped GaAs/AlGaAs heterostructures of
medium mobilities and rather high carrier concentrations, i.e.
with parameters typical for the samples recommended for
metrological applications of the IQHE. The current dependence of
their longitudinal resistivity at half-integer filling factors,
i.e. within the extended states in the middle of Shubnikov - de
Haas peaks in $\varrho_{xx}(B)$, has been described elsewhere
\cite{Nachtw1}.

\section{Experiments}
Two different samples have been employed in this study. Both were
made from wafers grown by MBE and patterned into Hall bar
geometry with 3 equidistant pairs of potential contacts separated
by L = 550 $\mu m$ along the channel of the width $w$. Sample A (referred to
as CS60-9 in ref. \cite{Nachtw1}) had following basic parameters :
$n_s(4.2K) = 5.4 \times 10^{15} m^{-2}$ ; $\mu (4.2K) = 39 T^{-1}$ ;
$w=400 \mu m$ . The other sample denoted here as B
(CS50-6 in ref. \cite{Nachtw1}) has been characterized by the
values : $n_s(4.2K)=3.8 \times 10^{15} m^{-2}$ ; $\mu (4.2K)=24
T^{-1}$ ;
$w=100 \mu m$ .

Both components of the resistivity tensor $\varrho_{xx}(T)$ and
$\varrho_{xy}(T)$ have been simultaneously measured at
temperatures between 4.2K and 85K using a computer controlled
data acquisition system with a voltage resolution better than 100
$n V$. The measuring d.c. current $I$ has been varied
within the interval $I = 1 - 100 \mu A$ and $I = 0.2 - 10 \mu A$ for
the sample A and B, respectively. Different currents used
for samples A and B correspond to their different widths $w$.
Measurement was performed in a
dynamical regime with the temperature continually changing in
both directions. The rate $dT/dt$ has been so slow, that no
hysteresis could by observed on the $\varrho_{xi}(T)$ (i=x,y)
curves throughout the whole temperature range. The conductivity
$\sigma_{xx}(T)$ was then calculated from the expression

\begin{equation}
\sigma_{xx}(T) = \frac{\varrho_{xx}(T)}{\varrho_{xx}^2(T) +
\varrho_{xy}^2(T)}
\end{equation}

\section{Results and discussion}
One of the remarkable features seen in both samples studied is a
pronounced asymmetry of the minimum in
$\varrho_{xx}(B)$ around $\nu=2$. This
asymmetry is moreover strongly dependent on current $I$, which is
illustrated in Fig.1a and Fig.1b for the samples A and B,
respectively. The resistivity minimum around $\nu$=2 is shown
here in the interval of currents covering the transition from
non-local to local resistivity discussed extensively elsewhere
\cite{Nachtw1,Svoboda}. While the upper (low-energy) edge of the
Shubnikov - deHaas (SdH) peak corresponding to the spin-resolved
Landau level 1$\uparrow$ are shown to depend strongly on the
current $I$ , this is not the case for the high-energy edge of the
peak 0$\downarrow$ on the high magnetic field side of the minimum.
The highest curves correspond to saturation
currents where there is a perfect coupling between edge and bulk
channels and further increase of $I$ leads to a suppression of
the heights of both SdH peaks 1$\uparrow$ (N=3) and
0$\downarrow$ (N=2) due to the
overheating of the electron gas \cite{Svoboda}.

An alternative explanation of this asymmetry
 of the line shapes of the individual spin-resolved subpeaks in
$\varrho_{xx}(B)$ for low-indexed Landau levels has been suggested
by Haug et al. \cite{Haug}.They have shown, that it can reflect
an asymmetry in the density of states induced by a particular
distribution of the attractive (ionized Si donors in AlGaAs
source layer) and repulsive (residual acceptors in GaAs buffer
layer) scatterers in samples. The situation observed here, where
the resistivity on the high-field side of the 1$\uparrow$ peak is
much more sensitive to both the current and the temperature than
the low-field side of the adjacent 0$\downarrow$ peak is then a result
of dominant contribution of remote attractive scatterers to the
conduction, which is typical for all heterostructures with
electron mobilities $\mu \geq 15 T^{-1}$.

We indicate in Fig.1 the range of magnetic fields $B_k$ in the
vicinity of the filling factor $\nu=2$, where the temperature
dependence of $\sigma_{xx}$ has been studied. All the fields
$B_k$ lie in the range where, within the resolution of our d.c.
method, $\varrho_{xx}(4.2K)=0$ and $\varrho_{xy}(4.2K)=h/2e^2$.
This resolution is limited mainly by the voltage noise and by the
stability of the current supply for the longitudinal and the Hall
resistivity, respectively, and it reaches about 0.1 $m \Omega$ at
highest currents used.

 Due to the asymmetry mentioned above, it is not
straightforward to find experimentally an exact center of the
IQHE plateaux, because it does not necessarily coincide with the
geometric center of the minimum in Fig.1, which is moreover
slightly current-dependent. As a criterion, we adopted the
temperature dependence of the Hall resistivity
$\varrho_{xy}(B_k,T)$, which is expected to be independent of T
at $\nu=2$. These dependences have been drawn in Fig.2 for two
different currents. It can be seen from Fig.2, that some degree of
quantization persists in the sample up to temperatures about 40K.

The temperature dependence of the conductivity $\sigma_{xx}$ in the
middle of the mobility gap is presented in the Arrhenius graph on
Fig.3 for $\nu \simeq$2 and $\nu \simeq$4, the only minima where at T=4.2K
the IQHE was still complete. Activated conductivity can be
detected below T$\simeq$20K. At $\nu=2.01$, the fitted activation
energy $\Delta = 9.22 meV$ ($\Delta/k_B = 107 K$) agrees well
with the value of $\hbar \omega_c /2 = 9.18 meV$ (106 K in
temperature units), which confirms a purely activated
conduction at low enough current $I$. The agreement is not so
good for $\nu = 4.08$ ($\Delta/k_B = 44.7K$ vers. $\hbar \omega_c
/2 k_B = 52.5K$), but it can be attributed to the fact, that we
are not just in the middle of the mobility gap.

Weiss et al. \cite{Weiss} suggested that a measurement of the
activated conductivity at various fixed magnetic fields within
the mobility gap can serve for a determination of the density of
(localized) states on the Fermi energy $D(E_F)$ provided that it
does not depend too strongly on the energy.
This method requires a measurement of $\sigma_{xx}(T,B=B_k)$ in a rather
dense set of precisely known fields $B_k$ around that
corresponding to an integer filling factor. Furthermore, it
works only within Landau level tails far from energies $E_N$.
Within the model of ref. \cite{Weiss}, the Fermi energy  shifts
due to a change of the field from $B_1$ to $B_2$ ($B_2 > B_1$) by

\begin{equation}\label{energie}
\delta E = \Delta(B_1) - \Delta(B_2) -
\frac{\hbar}{2}(\omega_{c,2} -
\omega_{c,1}).
\end{equation}

\noindent But the variation of the Fermi energy corresponds approximately
to a change of the carrier density

\begin{equation}\label{hustota}
\delta n \simeq \frac{\nu e}{2\pi \hbar}(B_2 - B_1).
\end{equation}

\noindent For $\nu = 2$, the density of states $D(E) = \delta n / \delta E$
 can then be
estimated from

\begin{equation}\label{def}
D(E) \simeq \frac{e}{\pi \hbar} \; \frac{(B_2 - B_1)}{[
\Delta(B_1) - \Delta(B_2)] - \frac{e \hbar}{2 m^*}(B_2 -
B_1)}.
\end{equation}

This approach assumes, that the mobility edge coincides with the
center of the Landau level $E_N$ and that it does not depend both
on the temperature and the carrier concentration. Although this
is probably oversimplified, we can at least roughly estimate from
the activation energies $\Delta(B_k)$ measured at low currents
near the middle of the mobility gap a mean
density of states. We have found that $D(E_F) \leq 2 \times 10^{10} meV^{-1} cm^{-2}$
within the interval of the halfwidth $\delta E \approx 2 meV$ around
the center of the mobility gap.
It should be compared with the zero-field value of
$D_o(E) = m^*/\pi \hbar^2 = 2.8 \times 10^{10} meV^{-1} cm^{-2}$.
Outside the above mentioned energy interval, the value of $D(E_F)$
increases and the increase is markedly steeper on the low-energy side,
i.e. at $\nu < 2$. Our samples thus apparently differ from the
higher mobility sample studied by Weiss et al. \cite{Weiss},
where a much lower and nearly constant $D(E_F)$ was reported over
a half of the mobility gap.

Fitting to the law (\ref{aktiv}) provides not only the energy
$\Delta$, but also the value of the pre-exponential factor
$\sigma_{xx}^o$. For the data presented in Fig.3 we get $\sigma_{xx}^o
=1.84 \times e^2/h$ and $\sigma_{xx}^o = 1.17 \times e^2/h$ for
$\nu \simeq$2 and $\nu \simeq$4 , respectively. The values of the prefactor
are however very sensitive to the exact position within the gap falling
steeply down with increasing $\mid \delta \nu \mid$.
This can be seen in Fig.4. Taking
the field $B_k$=10.68T as that corresponding to $\nu=2.00$ (see Fig.2),
we find that both curves in Fig.3 correspond to energies slightly
above the middle of the gap ($\nu=2.01$ and $\nu=4.08$, respectively).
Especially in the latter case it implies, that the value of $\sigma_{xx}^o$
at $\nu = 4$ has to be markedly higher than $e^2/h$. Our
data seems thus to be better explained by the theory of
Polyakov and Shklovskii \cite{Polyak2} for a long-range
scattering, predicting that $\sigma_{xx}^o \cong
2 e^2/h$. It is systematically higher both than the experimental
 result of Clark \cite{Clark}
$\sigma_{xx}^o \cong e^2/h$ and than the theoretical prediction
$\sigma_{xx}^o \leq e^2/h$ of Polyakov and Shklovskii \cite{Polyak1}
that should hold for samples with pure short-range scattering.

There is a clear difference among the curves presented in Fig.4.
While on the high-energy (low-field) side the activated
conductivity seems to persist even slightly outside the center,
 another contribution to the
conductivity appears on the low-energy side,
where the Fermi energy starts to move to the next fully occupied
Landau level 0$\downarrow$ . Only at $T > 10K$ we come back to the
common activated conductivity that persists then up to temperatures
 above  20 K.
As it is shown in Fig.5, qualitatively the same
contribution can be induced even in the very center of the gap by
increasing substantially the measuring current $I$.
Moreover, we demonstrate in Fig.6, that if this extra contribution to
$\sigma_{xx}(T)$ is already present at low currents and $\nu < 2$
it can be further enhanced by increasing the current $I$.

The most obvious explanation of such current-induced deviation
from the Arrhenius graph at lower temperatures could be given in
terms of an enhancement of the electron temperature above that of
the lattice, which we actually measure. To estimate this
contribution, we show in Fig.7 a part of the curves
$\varrho_{xx}(B)$ measured on sample A at several temperatures
between $T = 1.2K$ and $T = 4.2K$ with $I = 1 \mu A$ and compare
them with the data taken at $T = 1.2K$ using measuring currents
$I = 10 \mu A$ and $I = 50 \mu A$. We show here the minimum
centered at $\nu = 3$, where the Shubnikov- de Haas curves are
most sensitive to the temperature (there has been no measurable
temperature dependence of $\varrho_{xx}(B)$ at $\nu \simeq 4$).
 It can be seen from the graph,
that there is a measurable electron overheating at the current
$I = 50 \mu A$, but that it does not exceed (at least near the
center of the gap at $\nu \simeq 3$) about 0.3K at $T
\simeq 1.2K$. To explain the deviations from (\ref{aktiv})
observed in Fig.5 and 6. in terms of electron heating effects
only, one would need an enhancement of $T$ by an order of
magnitude higher. Moreover, it should occur at $T > 4.2K$, where
the transfer of thermal energy from electrons to the lattice
should be easier than at $T \simeq 1.2$. We can thus conclude,
that electron heating effects do not dominate in the data taken
with currents up to 50 $\mu A$.

An upward curvature from a linear Arrhenius graph at lower
temperatures has been observed in most studies of the activated
conduction in the IQHE regime. The onset temperature for this
deviation is apparently sample dependent. From the two commonly
used explanations, we can probably exclude a spurious influence of
non-ideal contacts \cite{Komiyama}. First, this kind of deviation
should lead to a saturation of $\sigma_{xx}(T)$ at a finite value
of the conductivity at
lowest temperatures, which is not seen in our samples at $T=4.2K$.
 Second, a measure
of the non-ideality of the contact is their resistance that has
been checked in our samples to be of the order of 10 $\Omega$ at
most, two orders of magnitude below those reported by Komiyama et
al. \cite{Komiyama}. The third argument stems from the microscopic
description of non-ideal contacts introduced in ref.
\cite{Komiyama} : Such contacts induce a non-equilibrium
population in the outgoing edge channels, that persists over
macroscopic distances along the sample. This non-equilibrium
between uppermost edge channels plays a crucial role in our model
explaining the current dependence of the amplitude of Shubnikov -
de Haas peaks in strong magnetic fields \cite{Svoboda}.
Sufficiently high currents  remove such a
non-equilibrium and possible influence of the contact should thus
be restricted to its immediate vicinity only. A contribution due
to contacts should thus be suppressed by a high current and just the
opposite can be seen in Fig.6.

It is widely accepted, that at low enough temperatures variable
range hopping (VRH), i.e. tunneling of the electrons from the interval
of the order of $k_B T$ among the localized states at $E_F$,
 prevails over the activated
conduction. The temperatures, where VRH in high mobility GaAs/AlGaAs
samples becomes dominant is expected to lie
at T$\leq$1K \cite{Ebert}. It has been stressed  by
Polyakov and Shklovskii \cite{Polyak1} that the actual form of
the $\sigma_{xx}(T)$ is governed by the overall character of the
disorder present in a particular sample. In the case of a purely
short-range scattering no dependence of the type (\ref{aktiv})
should be observed in any finite temperature interval
\cite{Polyak1}. Only an inflection point at $T=T_1$ is expected
to separate the VRH contribution at $T<T_1$ from the conduction
due to the broadened Fermi-Dirac distribution at $T>T_1$. It can
thus be expected, that the enhanced disorder in samples of a
lower quality would shift the interval of VRH conduction to
higher temperatures.

The contribution of the VRH in Landau level tails has been calculated
by Ono \cite{Ono} under the assumption that the magnetic field
causes a Gaussian localization of the electron wave function on
the scale given by the magnetic length
$\ell_c = \sqrt{\hbar/eB}$. His result reads

\begin{equation}\label{ono}
\sigma_{xx}^{VRH}(T) = \frac{e^2}{k_B T} \gamma_o
e^{-(T_o/T)^{1/2}}
\end{equation}

\noindent where $\gamma_o$ is a material parameter depending on
electron-phonon coupling and $T_o$ is a critical temperature
given by

\begin{equation}\label{to}
T_o = \frac{C}{k_B D(E_F) \ell_c^2}  \;\;\;  ;  \;\;\; C \approx 1
\end{equation}

It has been found that although the expression (\ref{ono}) fits
the data at lowest temperatures well, the density of states
calculated from the fitted critical temperatures $T_o$ on the
basis of (\ref{to}) becomes unrealistically high
\cite{Usher,Ebert}. Ebert et al. \cite{Ebert} reported values of
$D(E_F)$ by a factor 36 higher than the zero field density $D_o(E)
= m^*/\pi \hbar^2$.

The theory by Ono leading to eq. (\ref{ono}) assumes a finite
density of states at $E_F$ and unperturbed wave functions of
isolated impurities in the form
$\psi(r)\sim exp[-r^2/2 \ell_c^2]$. It has been criticised by
Polyakov and Shklovskii \cite{Shklov}, who provide another
expression for VRH conductivity in the form

\begin{equation}\label{ps}
\sigma_{xx}^{VRH}(T) = \sigma_o
e^{-(T_1/T)^{1/2}},
\end{equation}

\noindent which relies on the existence of a Coulomb gap at $E_F$
and assumes, that tails of the wave function have a simple
exponential form $\psi(r) \sim exp[-r/\xi]$ ($\xi$ being the
localization length) due to multiple
scattering of a tunneling electron \cite{Shklov}. The basic
difference between (\ref{ono}) and (\ref{ps}) is in the
expression  for the critical temperature; instead of (\ref{to})
they have got for $T_1$ the formula \cite{Shklov}

\begin{equation}\label{tjedna}
T_1(\nu) = C_1 \frac{e^2}{k_B \epsilon \xi(\nu)}
\end{equation}

\noindent with $\xi(\nu)$ denoting the localization radius of
the states on the Fermi energy for a given $\nu$, $\epsilon$ the
dielectric constant and with $C_1 \simeq 6$ in two dimensions.

Our data is not sufficient to distinguish between (\ref{ono}) and
(\ref{ps}). Both expressions are formally the same, because the
fitting of experimental data to (\ref{ps}) by Koch et al.
\cite{Koch} gives $\sigma_o \sim 1/T$. But while the fitting of our
data to (\ref{ono}) provides us with $T_o$ leading to reasonable
values of $D(E_F)$ (see below), critical temperatures $T_1$ obtained from
fitting the same data to (\ref{ps}) are by at least one order of
magnitude higher than those discussed by Polyakov and Shklovskii
\cite{Shklov}. We suggest, that the formula (\ref{ono}) is
therefore more relevant to our samples.

We present in Fig.8 the data measured on the
sample A with the saturation current
$I = 50\mu A$ \cite{Nachtw1} at a few magnetic fields $B_k$  around $\nu=2$. At
$T \leq 12 K$ the data can be fitted to the  formula
(\ref{ono}). This has not to be a convincing argument (we have
shown in Fig.5a, that the same data can be nearly equally well fitted
to the Arrhenius law (\ref{aktiv})), let us mention, however,
that the temperatures $T_o$ obtained give  very
reasonable values of the density $D(E_F)$ in our case . From the curves for
$B_k$=10.63T and $B_k$=10.73T, that correspond to an immediate
vicinity of the center $\nu=2$ we obtain $D(E_F) = 1.0 \times
10^{10} meV^{-1} cm^{-2}$ and $D(E_F) = 1.4 \times
10^{10} meV^{-1} cm^{-2}$ , respectively. These values agree with
those determined from the activated conductivity by the method of
Weiss et al. \cite{Weiss} and also a steep increase of $D(E_F)$
further from the center is similar here. The values of the
pre-exponential factor $\sigma_{xx}^* = e^2 \gamma_o /k_B $
strongly decrease if the field $B_k$ is shifted
with respect to this corresponding to $\nu=2$.
It is hard to understand why the quantity $\gamma_o$ being a
material parameter which should be a function
 of the electron-phonon coupling
strength only \cite{Ono} would depend so strongly on the field $B_k$
and/or on the density of states $D(E_F)$. The corresponding
changes of the filling factor are so small, that any pronounced
changes of the density of states at $E_{F}$ can hardly be
expected.

Even the conductivity measured by a low current $I$ at $\nu < 2$ can be
at $T \leq 12K$ reasonably well approximated by formula (\ref{ono}), as it
can be seen from Fig.9. An enhancement of the current $I$ without changing
the filling factor results there in an apparent suppression of the critical
temperature $T_o$ which means in terms of (\ref{to}) an enhancement of the
density $D(E_F)$. This can be understood since high measuring currents lead
to a strong electric (Hall) field across the sample which may
result in an additional
broadening of the Landau levels.

As we have mentioned already , the extra
contribution to the activated conductivity
(but not the activated conductivity itself) can be
fitted both to (\ref{ono}) (see Fig.5) and (with nearly the same
accuracy) to the expression (\ref{aktiv}) . It is
however hard to interpret the activation energy $\Delta^*
\approx \Delta/2$ obtained in the
latter case. Such a fit gives also very low values of the
pre-exponential factor $\sigma_{xx}^o \approx 0.01 \times e^2/h$,
which
are not compatible with any of the existing theoretical descriptions
of the activated conduction.
The fact,
that our data can be fitted to an expression of the type (\ref{ono})
therefore indicates, that a contribution of hopping among the localized
states around the middle of the mobility gap persists to much higher
temperatures than those reported before \cite{Usher,Ebert}.

In the B\"{u}ttiker - Landauer picture of the IQHE, a finite
longitudinal conductivity $\sigma_{xx}$ appears once a
backscattering between the two edges of the sample occurs. This
happens if there is a mechanism capable to transfer an electron
injected by one current contact into the edge channels on one
side of the sample across the bulk composed of localized states
only into the edge channels on the opposite side before it can
arrive into the other current contact. At low measuring currents
a gradient of electrochemical potential starts to develop near
the edges without influencing the potential far from the edge
\cite{Nachtw2}. If, however, the current exceeds some (sample
dependent) critical value, a finite gradient of the
electrochemical potential develops throughout the sample.
The potential distribution across an IQHE sample at low and high
sample currents has been recently calculated by Cage and Lavine
\cite{Cage}. They found that at high currents $I$ a so-called
charge-redistribution potential $V_r(y)$ adds to common confining
potential $V_c(y)$ that gives rise to the edge channels. Far from
the edges, $V_r(y)$ changes linearly with the coordinate $y$ and
its slope is proportional to $I$. The charge-redistribution potential
$V_r(y)$ should provide a "driving force" for the transfer of the charge
across the bulk of the sample.

We suggest that the enhanced contribution to the conductivity beyond the
temperature induced breakdown of the IQHE observed near the middle
of the mobility gap around $\nu=2$ can be due to enhanced tunneling
through the potential barriers separating localized states at $E_F$.
It can be connected both with the thermally assisted tunneling if the
density of localized states is sufficiently high and with  the
transversal electric (Hall) voltage $U_H^b \sim I$ developing within the
bulk of the sample with a
finite conductivity $\sigma_{xx}(T)$.
 The latter gives rise to an electric field
assisted tunneling,
which further promotes backscattering at higher measuring currents $I$.
Due to this enhanced tunneling, the temperature dependence of the
conductivity contains a contribution of the hopping conductivity up
to temperatures $T \simeq 12K$. This temperature corresponds to a
thermal
energy of 1 meV ,which is about 10\% of the halfwidth of the mobility
gap in our experimental condition.
At still higher temperatures, activated conductivity starts to prevail
again and it could be observed up to $T \simeq 25K$ where the Boltzmann
statistics can not be used any more and a downward deviation from
the Arrhenius graph starts to develop.

\section{Conclusions}
The temperature dependence of the longitudinal conductivity
$\sigma_{xx}$ has been studied on samples with well developed
IQHE plateaux in magnetic fields around the centers of highest
mobility gaps corresponding to filling factors $\nu=2$ and
$\nu=4$. At low measuring currents $I$ pure activated
conductivity described by the expression (\ref{aktiv}) has been
observed in the center of the mobility gaps. The
activation energies $\Delta$ obtained by fitting the data
to the Arrhenius law (\ref{aktiv}) scale well with the
half-width $\hbar \omega_c/2$ of the interval between adjacent
Landau levels. The prefactor $\sigma_{xx}^o$ in (\ref{aktiv}) depends
markedly on the exact position of the Fermi level $E_F$ with
respect to the center of the mobility gap. At integer filling
factors it approaches the value of about $2 e^2/h$. This
value is consistent with
the theoretical predictions of Polyakov and Shklovskii
\cite{Polyak2} for samples with dominating long-range scattering,
but it does neither support the experimental findings
of Clark \cite{Clark} nor the prediction
$\sigma_{xx}^o \leq e^2/h$ calculated for a pure short-range scattering
\cite{Polyak1}.

Further from the center of the mobility gap, another contribution
to $\sigma_{xx}(T)$ develops at $T \leq 12K$. At low enough currents,
this contribution can be seen on the high-field (low-energy) side
of the center only, where the density of localized states
increases more steeply than on the low-field side. This
asymmetry of $D(E)$ can be of the same origin as that
reported by Haug et al. \cite{Haug}.

This extra contribution can be induced even in the middle of the
mobility gap by increasing the current $I$ up to the values where
the non-local conduction \cite{Nachtw1,Svoboda} is suppressed and
decoupling between the edge and bulk channels has been removed.
From the temperature dependence of $\varrho_{xx}(B)$ measured at
$T \leq 4.2K$ with various currents we can conclude, that this is
not a simple electron overheating effect.

The additional contribution to $\sigma_{xx}(T)$ can be formally fitted
both to the expression for the activated conduction (\ref{aktiv})
and to the formula (\ref{ono}) derived for the variable range
hopping in strong magnetic fields.
 In the first case we obtain activation energies
$\Delta^*$ that do not scale with the separation of adjacent
Landau levels and that are substantially smaller than "true"
activation energies $\Delta$.
Pre-exponential factors
$\sigma_{xx}^o$ in (\ref{aktiv}) become very small, well below
any theoretical prediction for the activated conductivity.

Fitting to (\ref{ono}) gives
densities of states $D(E_F)$ calculated from
(\ref{to}) that agree well with those estimated
from the activation energies $\Delta$ according to ref.
\cite{Weiss}. Both methods give for the density just in the
middle of the mobility gap values $D(E_F) \approx (1-2) \times
10^{10} meV^{-1} cm^{-2}$, which seems to be rather high but not
unreasonable if compared with the zero-field density $D_o(E) =
2.8 \times 10^{10} meV^{-1} cm^{-2}$.

We suggest that the additional contribution to $\sigma_{xx}(T)$
can be a result of an enhanced backscattering of electrons from
one edge of the sample to the other one.
The fact, that the data
can be reasonably well fitted to an expression derived for
the variable range hopping in strong magnetic fields, seems to
indicate that it is the tunneling through the potential barriers separating
localized states at $E_F$ that contributes to the backscattering.
At high enough densities of the localized states $D(E_F)$ such a tunneling
can apparently lead to a hopping conductivity that can be seen even
at temperatures well above 4.2K. In addition to thermally assisted
tunneling, the current dependence of our data indicates that it can
be an electric field assisted tunneling driven by the transversal
electric field
arising  due to the Hall voltage
developing across the bulk of the sample once its conductivity becomes
finite.

Upon increasing the current $I$ injected into the sample, an
accumulation of the charge in the edge channels gives rise to a
charge-redistribution potential \cite{Cage} that adds to the
common confining potential. At high enough currents, a nearly
constant potential gradient develops across the sample. We
believe that this gradient further enhances the probability of
crossing the potential barriers within the bulk. It supports the
backscattering between the edges which manifests itself in an
enhanced longitudinal conductivity or resistivity of the sample
at higher currents.

\vspace{10mm}
\Large
{\bf Acknowledgement}

\vspace{3mm}
\normalsize
The authors wish to thank Prof. B. I. Shklovskii for his remarks
and comments.
This work has been supported by the Grant Agency of the ASCR
under the contract No. 110423. P.S. acknowledges financial
support from the Deutsche Forschungsgemeinschaft (DFG) during his
stay at the Humboldt-Universit\"{a}t zu Berlin.

\newpage

\Large
{\bf Figure captions}

\vspace{10mm}
\normalsize

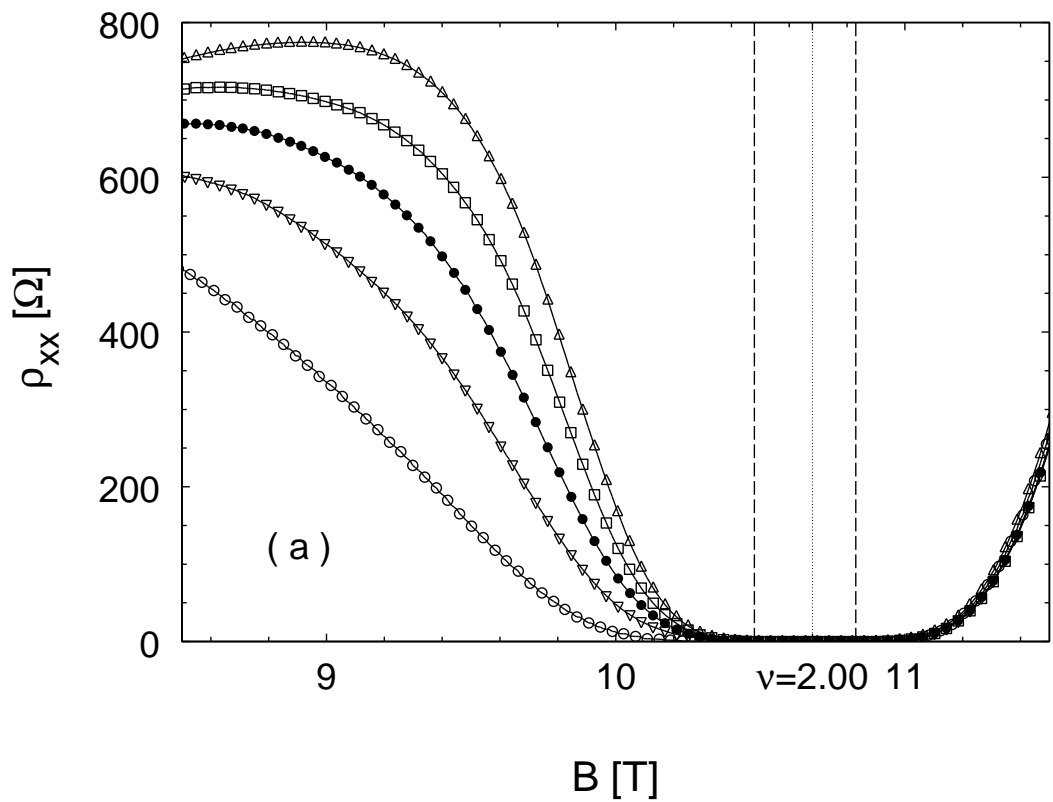
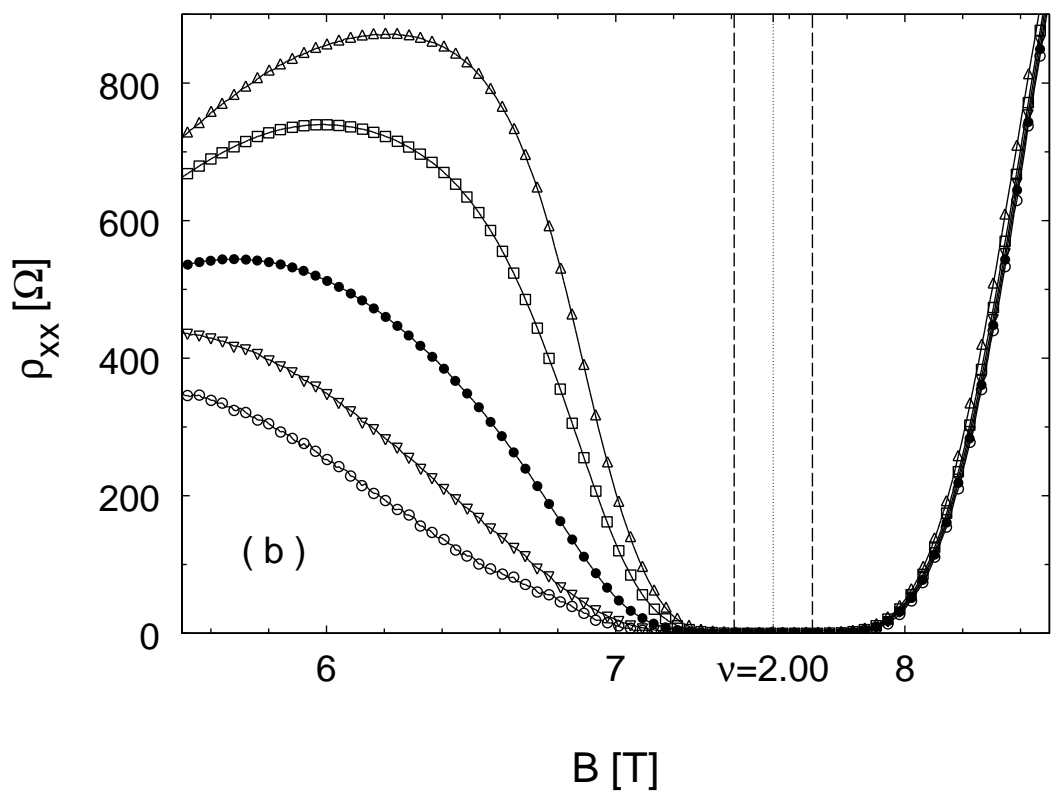
{\bf Fig. 1 :} Longitudinal resistivity $\varrho_{xx}(B)$
at $T = 4.2K$ for the sample
A {\bf (a)} and B {\bf (b)} in the vicinity of the minimum corresponding to the filling
factors $\nu \approx 2$. Measuring currents $I$ cover the interval
where the transition from non-local to local conduction [9,10]
occur for these samples.
The curves correspond to following currents : {\bf (a)} .... $I=
1 \mu A \;$ ({\Large $\circ$}); $I= 5 \mu A  \; (\nabla)$;
$I=10 \mu A \;$ ({\Large $\bullet$}); $I=20 \mu A \; (\Box)$;
$I=50 \mu A \; (\triangle)$. {\bf (b)} .... $I=
0.2 \mu A \;$ ({\Large $\circ$}); $I= 1 \mu A \; (\nabla)$;
$I= 2 \mu A \;$ ({\Large $\bullet$}); $I= 5 \mu A \; (\Box)$;
$I=10 \mu A \; (\triangle)$.
 Vertical dashed lines
indicate the range of magnetic fields $B_k$ around $\nu=2.00$
 where the temperature dependence of the resistivity $\varrho_{xx}(T)$ and
 $\varrho_{xy}(T)$ has been studied.

\vspace{5mm}

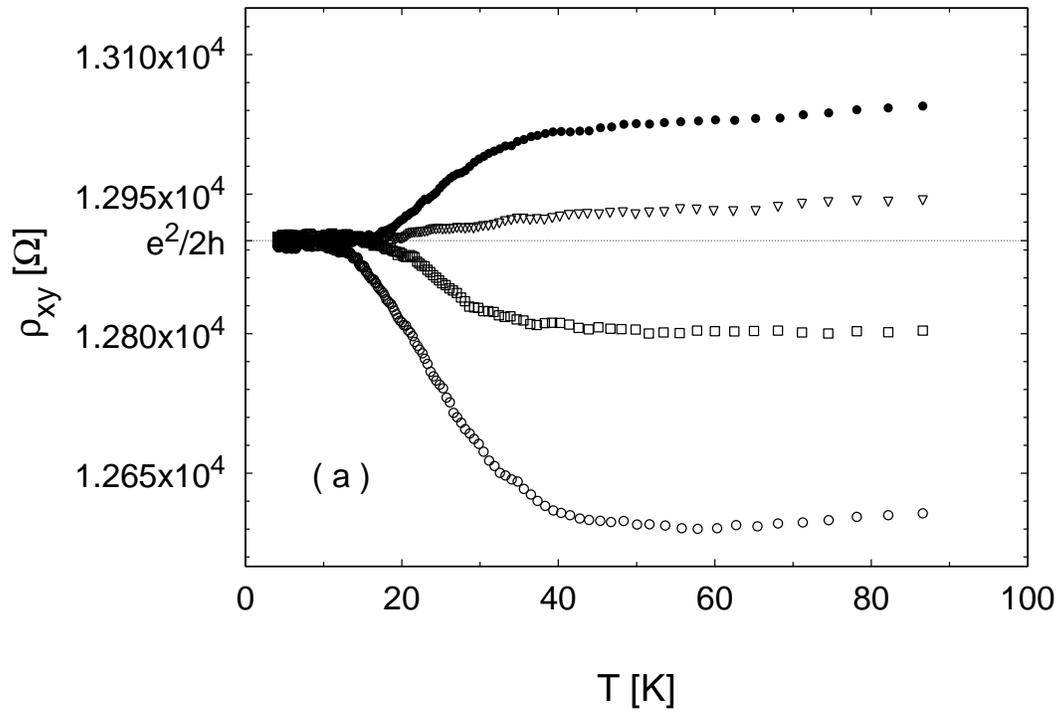
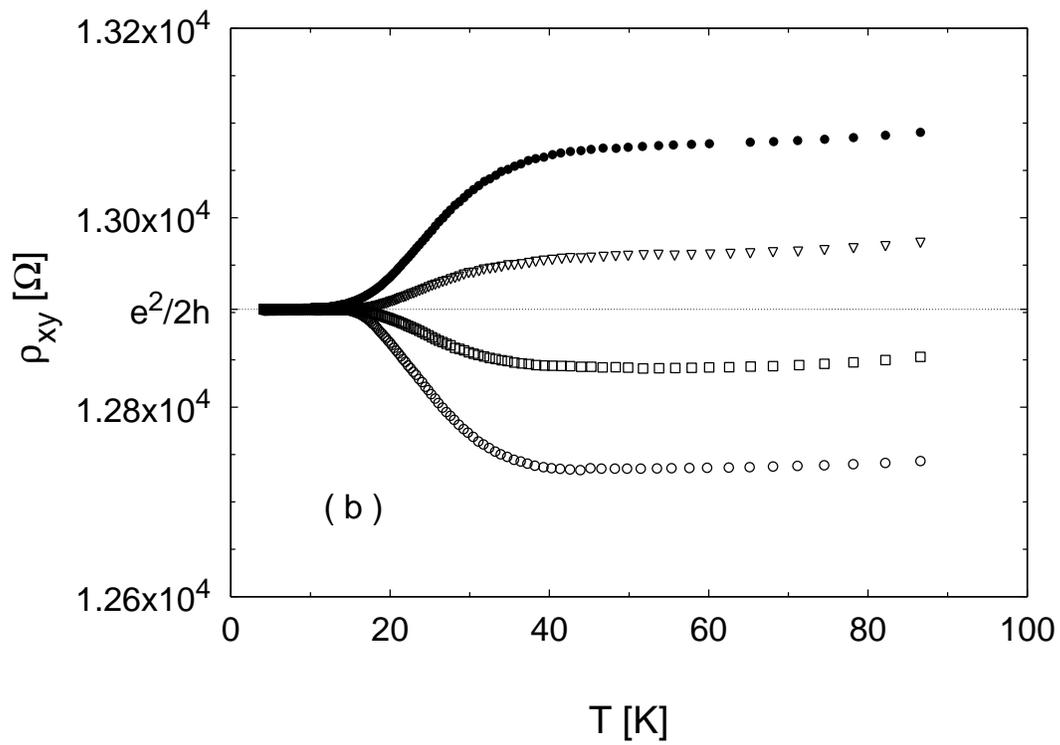
{\bf Fig. 2 :} Temperature dependence of the Hall resistivity
$\varrho_{xy}(T)$  in the middle of the second IQHE plateaux
measured on the sample A.
Just in the center of the plateaux
Hall resistivity should be independent of the temperature and given
by $\varrho_{xy} = h / 2 e^2$.

\noindent {\bf ( a )} ..... $I= 1 \mu A$ ; {\Large $\circ$}...
$B_1=10.48T \;$
 ($\nu=2.04$), $\Box$... $B_2=10.63T \; (\nu=2.01)$, $\nabla$...
$B_3=10.73T \; (\nu=1.99)$, {\Large $\bullet$}... $B_4=10.83T \;
(\nu=1.97)$.

\noindent {\bf ( b )} ..... $I=50 \mu A$ ; {\Large $\circ$}...
$B_1=10.53T \;$
 ($\nu=2.03$), $\Box$... $B_2=10.63T \; (\nu=2.01)$, $\nabla$...
$B_3=10.73T \; (\nu=1.99)$, {\Large $\bullet$}... $B_4=10.83T \;
(\nu=1.97)$.

\vspace{5mm}

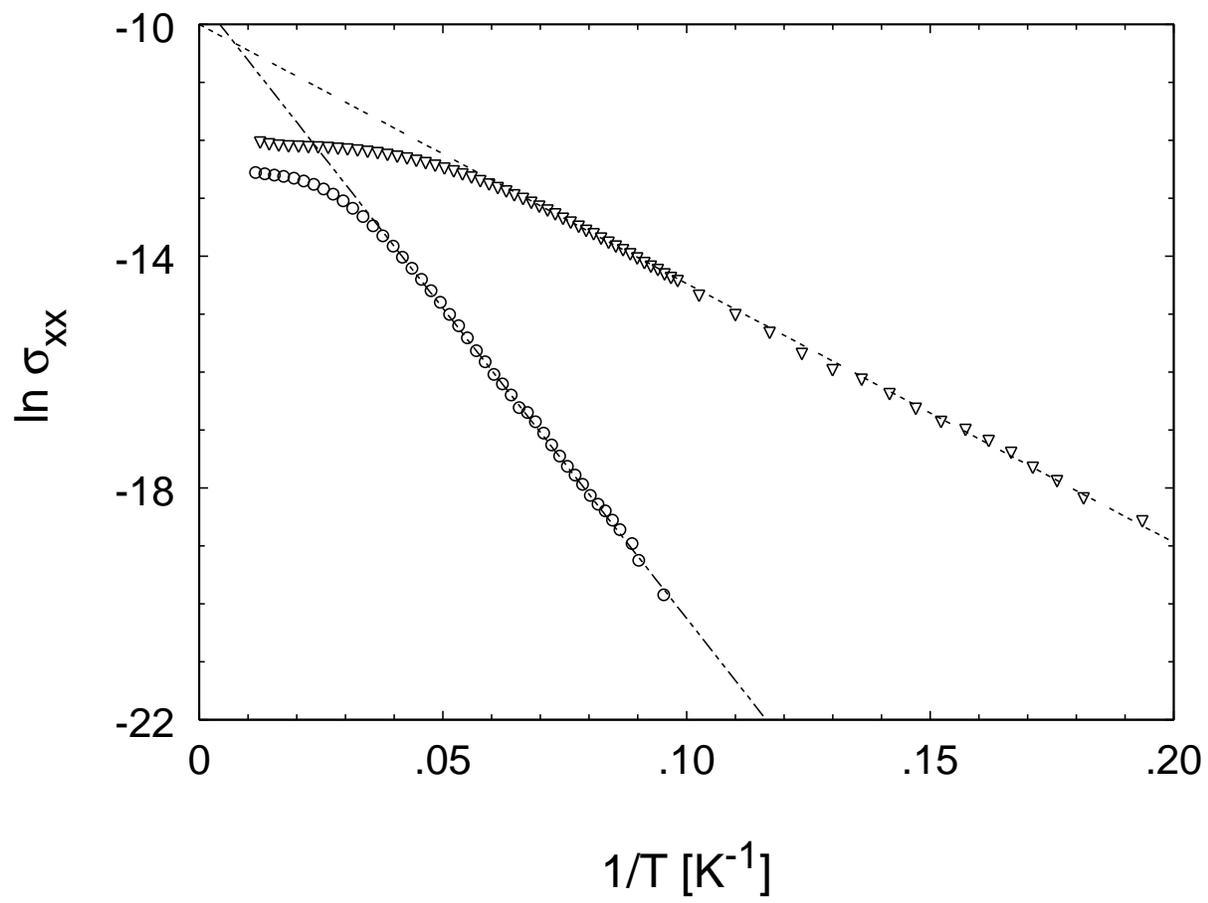
{\bf Fig. 3 :} Temperature dependence  of the longitudinal conductivity
$\sigma_{xx}(T)$ for the sample A
drawn in the Arrhenius graph for the fields $B = 10.63 T$
({\Large $\circ$}) and $B = 5.24 T$
$ (\nabla)$, corresponding to filling factors $\nu = 2.01$ and
$\nu = 4.08$, respectively. Straight lines
in the graph indicate the best fit to formula for activated
conductivity (1). The parameters obtained by the fitting are
:

\noindent $\sigma_{xx}^o = 1.84 e^2/h ; \Delta/k_B = 107 K$  and
$\sigma_{xx}^o = 1.17 e^2/h ; \Delta/k_B = 44.7 K$ for  $\nu=2.01$ and
$\nu=4.08$,
respectively.

\vspace{5mm}

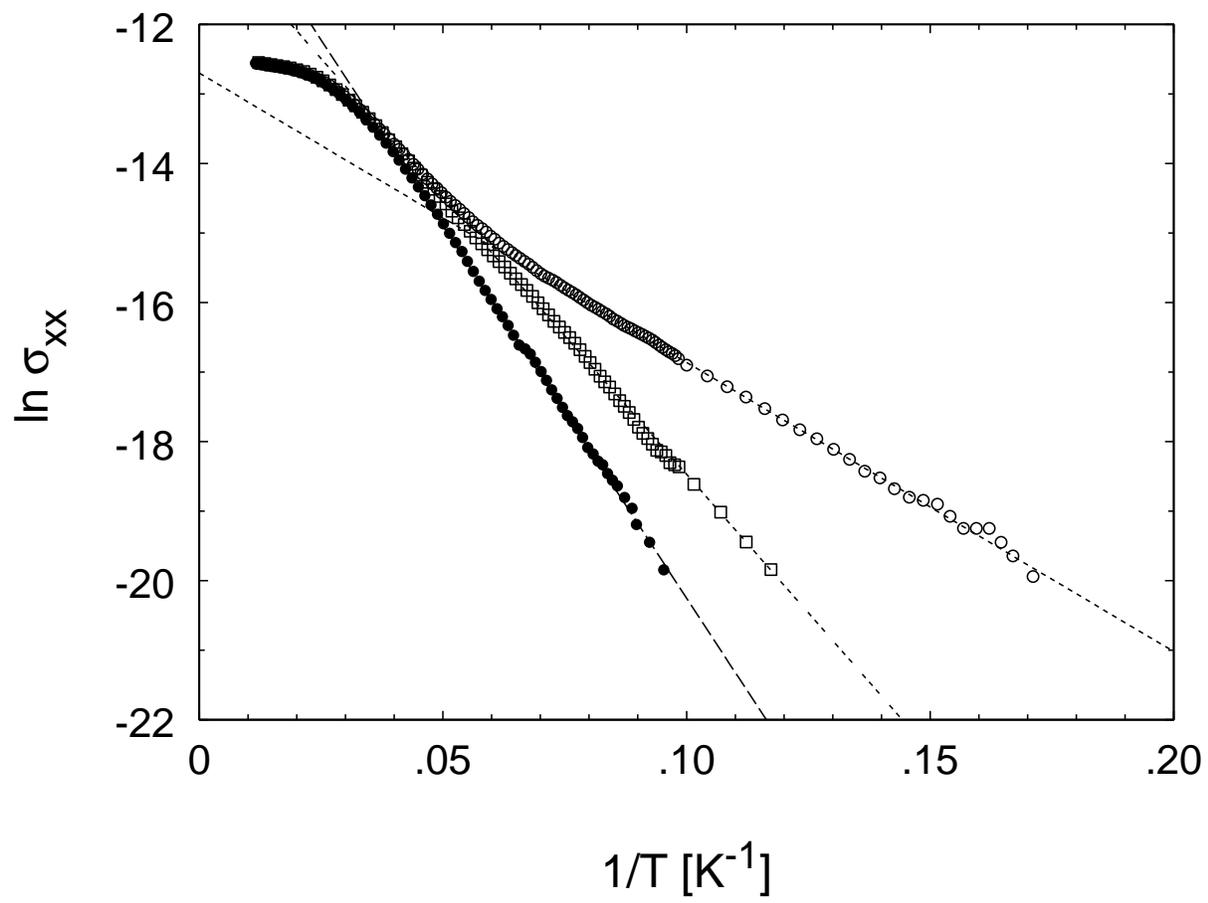
{\bf Fig. 4 :} Temperature dependence of the longitudinal conductivity
$\sigma_{xx}(T)$ for sample A  measured with $I=1\mu A$
on both sides of the center
of the mobility gap at $\nu=2$ and fitted to the expression (1).
$\Box$ ..... $\nu=2.04$, $\sigma_{xx}^o = 0.71 e^2/h$, $\Delta/k_B = 79.7 K$;
{\Large $\bullet$} .....
 $\nu=2.01$, $\sigma_{xx}^o = 1.84 e^2/h$, $\Delta/k_B = 107 K$;
{\Large $\circ$} .....
  $\nu=1.97$, $\sigma_{xx}^o = 0.08 e^2/h$, $\Delta/k_B = 41.6 K$
  ( for $T < 12 K$ ).

\vspace{5mm}

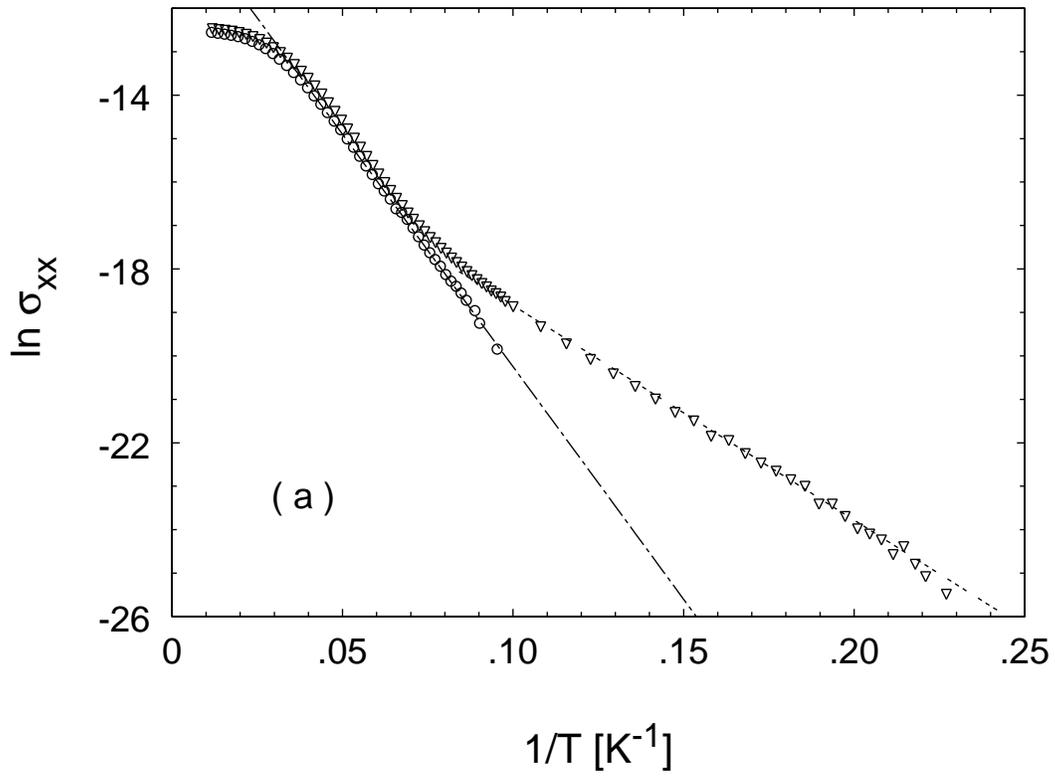
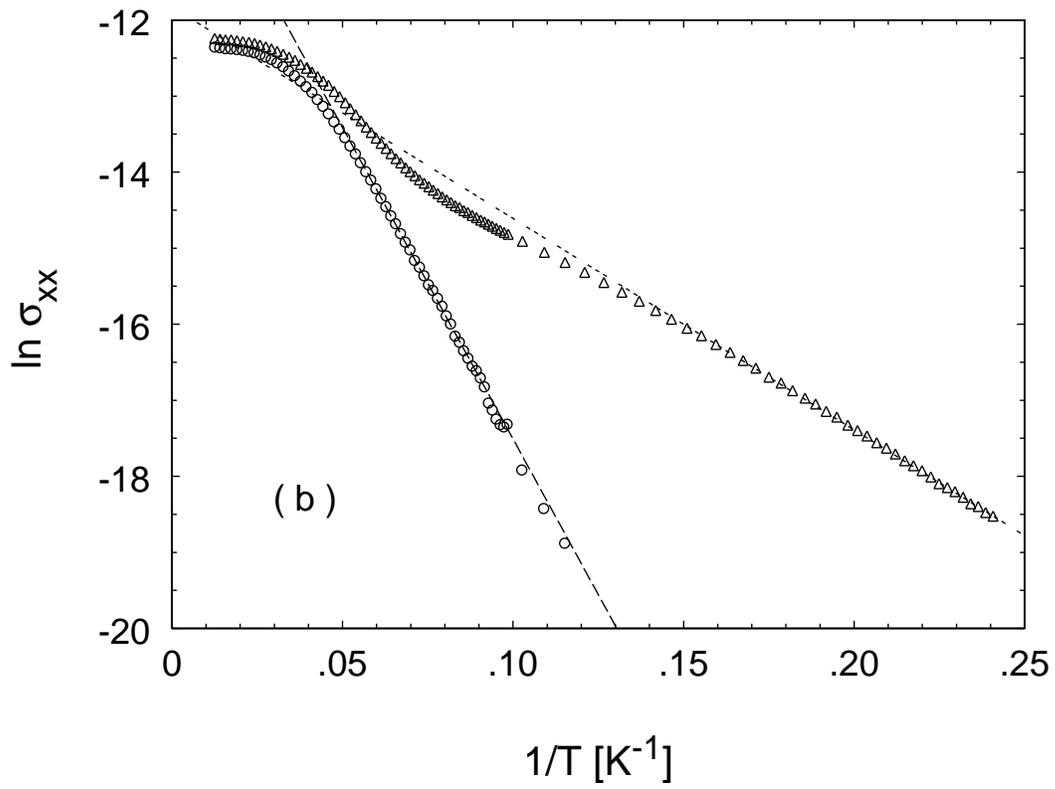
{\bf Fig. 5 :} Conductivity $\sigma_{xx}(T)$  measured in the center
of the mobility gap at $\nu=2$ using various currents $I$ and
fitted to (1).

\noindent {\bf (a)} .... sample A : {\Large $\circ$} ...
 $I = 1 \mu A$; $\sigma_{xx}^o = 1.84 e^2/h$,
 $\Delta/k_B = 107 K$;  $\nabla$ ...
 $I = 50 \mu A$ , $\sigma_{xx}^o = 0.02 e^2/h$,
 $\Delta/k_B = 49.7 K$ (for $T < 12 K$) ;

\noindent {\bf (b)} .... sample B : {\Large $\circ$} ... $I = 0.2 \mu
A$;
 $\sigma_{xx}^o = 2.31 e^2/h$,
 $\Delta/k_B = 82 K$;  $\triangle$ ...
 $I = 10 \mu A$; $\sigma_{xx}^o = 0.2 e^2/h$,
 $\Delta/k_B = 28 K$ ( for $T < 7 K$ ).

\vspace{5mm}

\newpage

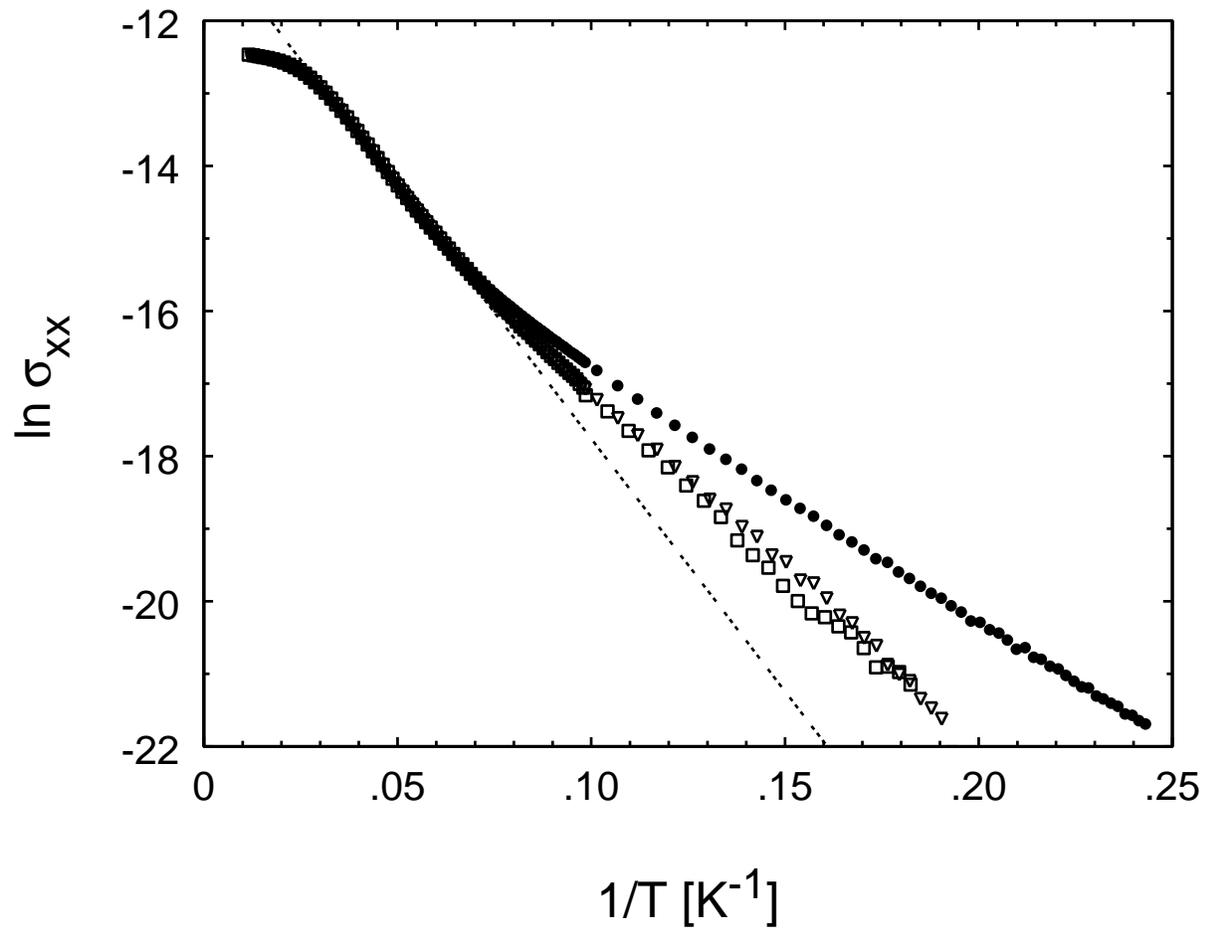
{\bf Fig. 6 :}  Conductivity $\sigma_{xx}(T)$  of the sample A at
$\nu=1.98$ for the currents $I=1\mu A$ ($\Box$), $I=10\mu A$
 ($\nabla$) and $I=50\mu A$ ({\Large $\bullet$}).
The dashed line represents a linear extrapolation
of the activated conductivity at $T \geq 12 K$.

\vspace{5mm}

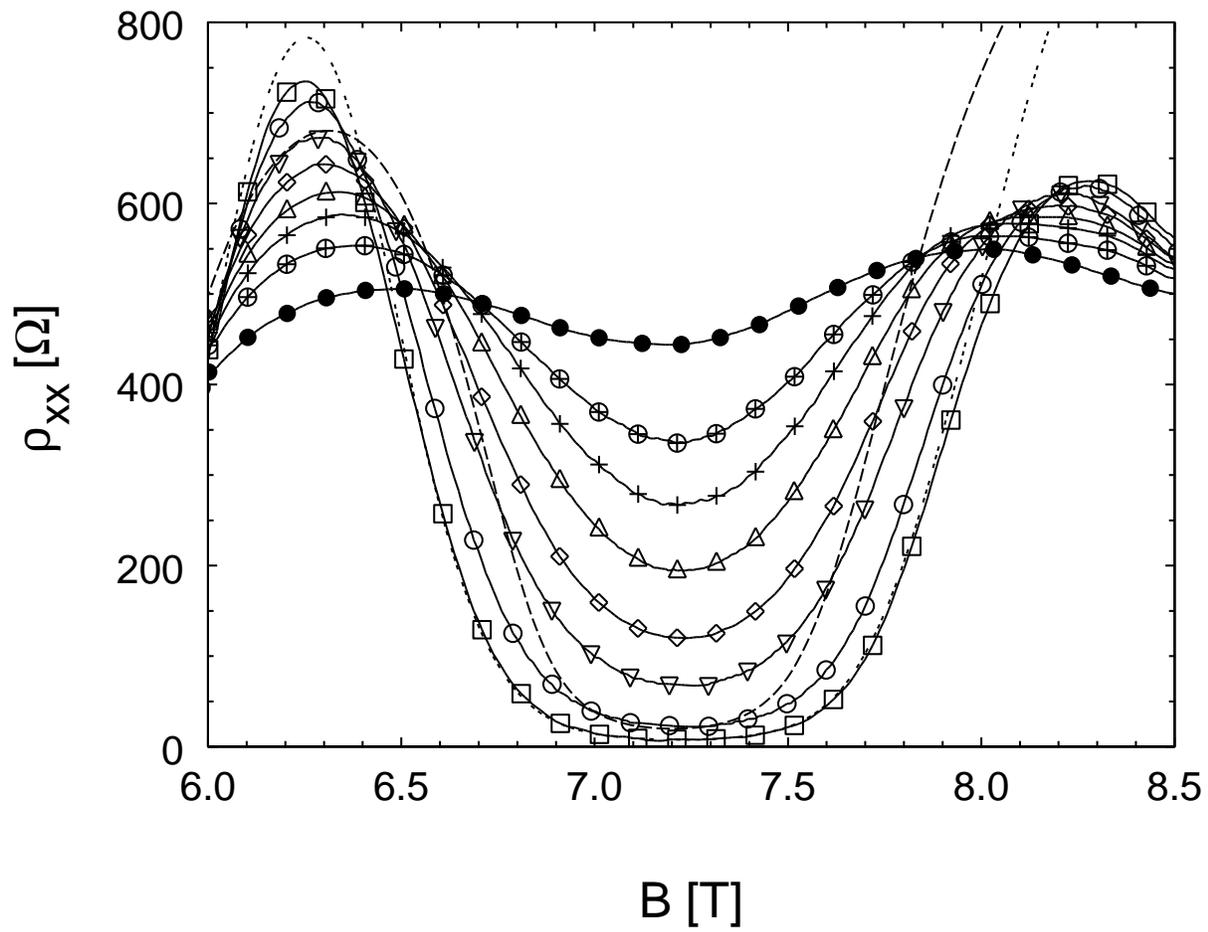
{\bf Fig. 7 :} Longitudinal resistivity of the sample A near the
minimum corresponding to $\nu=3$. Full lines correspond to the
current $I = 1 \mu A$ and to the temperatures $T=1.23K (\Box)$,
$T=1.45K$ ({\Large$\circ$}), $T=1.81K (\nabla)$, $T=2.11K
(\Diamond)$, $T=2.42K (\triangle)$, $T=2.75K$ ({\Large +}),
$T=3.11K (\bigoplus)$ and $T=4.23K$ ({\Large $\bullet$}).
Dotted line is for $T=1.23K$ and $I=10 \mu A$, the dashed
line for the same temperature but $I=50 \mu A$.

\vspace{5mm}

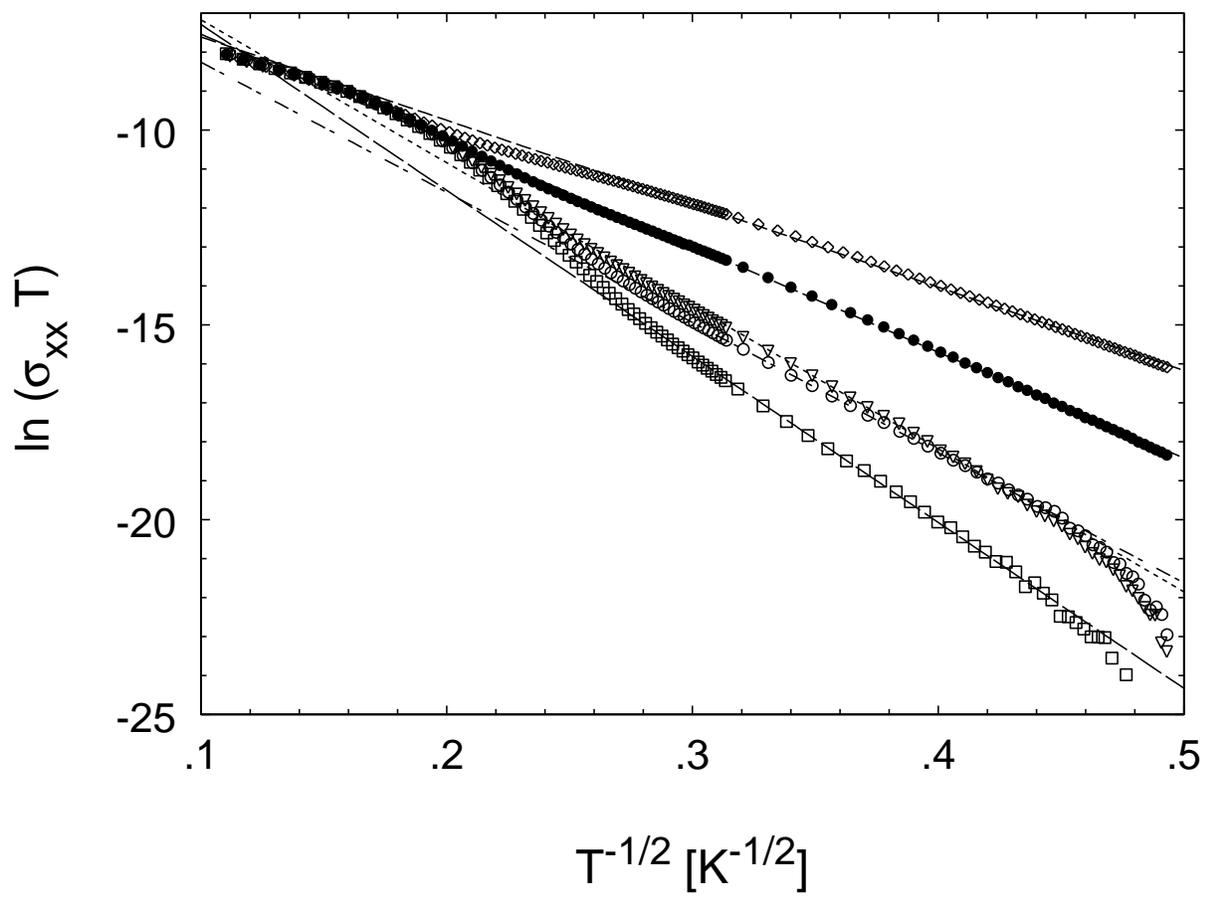
{\bf Fig. 8 :} Fitting of the data on  $\sigma_{xx}(T)$ measured in sample
A using $I=50\mu A$ at various filling factors  to the expression
(5) with $\sigma_{xx}^* = e^2 \gamma_o / k_B$.

\noindent  {\Large $\circ$} ... $\nu=2.03$; $\sigma_{xx}^* = 0.007$,
 $T_o=1116K$;
$\Box$ ... $\nu=2.01$; $\sigma_{xx}^* = 0.048$, $T_o=1815K$;
$\nabla$ ... $\nu=1.99$; $\sigma_{xx}^* = 0.030$, $T_o=1347K$;
{\Large $\bullet$} ... $\nu=1.97$; $\sigma_{xx}^* = 0.008$, $T_o= 740K$;
$\diamondsuit$ ... $\nu=1.95$; $\sigma_{xx}^* = 0.004$, $T_o=
458K$.

\vspace{5mm}

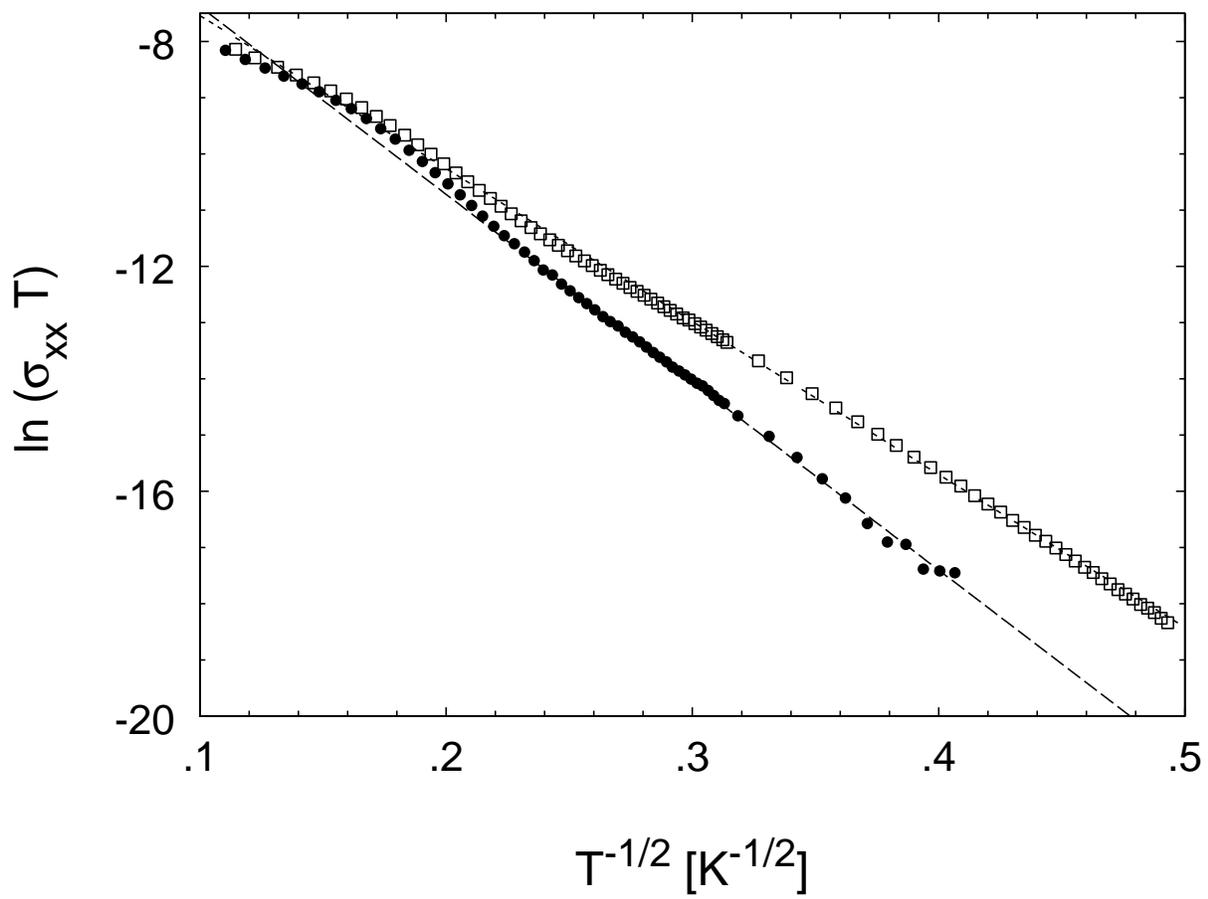
{\bf Fig. 9 :} Fitting of the data on  $\sigma_{xx}(T)$ for the sample
A at $\nu=1.97$ to the expression (5). The parameters of the fit
at temperatures $T \leq 12 K$ are :
{\Large $\bullet$} .....
 $I= 1\mu A$; $\sigma_{xx}^* = 0.018$, $T_o=1116K$;
$\Box$ ..... $I=50\mu A$; $\sigma_{xx}^* = 0.008$, $T_o= 740K$.


\newpage
\begin{thebibliography}{16}

\bibitem{Clark} R.G.Clark 1991 {\it Physica Scripta} {\bf T39}  45
\bibitem{Usher} A. Usher, R.J. Nicholas, J.J. Harris and C.T.
Foxon 1990 {\it Phys. Rev.} {\bf B 41} 1129
\bibitem{Weiss} D.Weiss, K. von Klitzing and V. Mosser 1986 {\it in
Springer Series in Sol. St. Phys., Vol. 67 } eds. G. Bauer {\it
et al.}
(Springer Verlag) pp. 204-217
\bibitem{Katayama} Y. Katayama, D.C. Tsui and M. Shaygan 1994 {\it
Phys. Rev.} {\bf B 49} 7400
\bibitem{Polyak1} D.G. Polyakov and B.I. Shklovskii 1994 {\it Phys.
Rev. Lett.} {\bf 73} 1150
\bibitem{Polyak2} D.G. Polyakov and B.I. Shklovskii 1995 {\it Phys.
Rev. Lett.} {\bf 74}, 150
\bibitem{Ebert} G. Ebert, K. von Klitzing, C. Probst, E.
Schuberth, K. Ploog and G.Weimann 1983 {\it Sol. St. Commun.} {\bf
45} 625
\bibitem{Komiyama} S. Komiyama, H. Hirai, S. Sasa and T. Fujii
1990
{\it Sol. St. Commun.} {\bf 73} 91
\bibitem{Nachtw1} G. Nachtwei, C. Breitlow, J. Sayfarth, S.
Heide, L. Bliek, F.-J. Ahlers, P.Svoboda and M. Cukr 1994 {\it
Semicond. Sci. Technol.} {\bf 9} 10
\bibitem{Svoboda} P. Svoboda, P. St\v{r}eda, G. Nachtwei, A.
Jaeger, M. Cukr and M. L\'{a}zni\v{c}ka 1992 {\it Phys. Rev.} {\bf B
45} 8763
\bibitem{Haug} R.J. Haug, K. von Klitzing and K. Ploog 1987 {\it
Phys. Rev.} {\bf B 35}, 5933
\bibitem{Ono} Y. Ono 1982 {\it J. Phys. Soc. Japan} {\bf 51} 237
\bibitem{Shklov} D.G.Polyakov and B.I.Shklovskii 1993 {\it Phys.
Rev.} {\bf B 48}, 11167
\bibitem{Koch} S.Koch, R.J.Haug, K.von Klitzing and K.Ploog 1995
{\it Semicond. Sci. Technol.}{\bf 10}, 209
\bibitem{Nachtw2} G. Nachtwei, S. Heide, C. Breitlow, P. Svoboda
and M. Cukr 1994 {\it Phys. Rev.} {\bf B 50} 8488
\bibitem{Cage} M.E. Cage and C.F. Lavine 1995 {\it J. Res. Nat.
Stand. Tech.} {\bf 100} 529

\end{thebibliography}
\end{document}